\documentclass[aps,prb,reprint,noeprint]{revtex4-2}

\usepackage{epsfig}
\usepackage{graphicx}
\usepackage[figuresright]{rotating}

\usepackage{amsfonts}
\usepackage{amsmath}
\usepackage{amssymb}
\usepackage{wasysym}
\PassOptionsToPackage{normalem}{ulem}
\usepackage{ulem}
\usepackage[unicode=true,pdfusetitle,
 bookmarks=true,bookmarksnumbered=false,bookmarksopen=false,
 breaklinks=false,pdfborder={0 0 1},backref=false,colorlinks=false]
 {hyperref}
\hypersetup{colorlinks,linkcolor=blue,citecolor=blue,urlcolor=blue}

\makeatletter

\usepackage{xcolor}
\usepackage{pdfcolmk}
\usepackage{wasysym}

\usepackage{txfonts}
 
\makeatother

\hypersetup{
	pdfborder={0 0 0},
	pdfnewwindow=true,      
	colorlinks=true,       
	linkcolor=blue,          
	citecolor=blue,        
	filecolor=blue,      
	urlcolor=blue,          
}
\usepackage{graphicx} 
\usepackage{amsmath} 
\usepackage[version=4]{mhchem} 

\newcommand{\Fig}{Fig.~}

\renewcommand{\vec}[1]{\boldsymbol{#1}}
\renewcommand{\tensor}[1]{\stackrel{\leftrightarrow}{\vec{#1}}}
\newcommand{\me}{e}
\newcommand{\mi}{\imath}

\usepackage{ulem} 
\usepackage{cancel} 

\newcommand{\del}[1]{\ifmmode \textcolor{red}{\xcancel{#1}} \else \textcolor{red}{\sout{#1}} \fi} 
\newcommand{\rev}[2]{\ifmmode \textcolor{red}{\xcancel{#1}} \textcolor{red}{#2} 
	\else \textcolor{red}{\sout{#1}} \textcolor{red}{#2} \fi} 

\newif\ifdraft
\draftfalse

\newif{\ifshowcomments}
\showcommentstrue

\begin{document}
	
\title{Twisted magnetic topological insulators}

\author{Chao-Kai Li}
\thanks{These authors contributed equally to this work.}
\affiliation{Department of Physics and HKU-UCAS Joint Institute 
for Theoretical and Computational Physics at Hong Kong, 
The University of Hong Kong, Hong Kong, China}
\author{Xu-Ping Yao}
\thanks{These authors contributed equally to this work.}
\affiliation{Department of Physics and HKU-UCAS Joint Institute 
for Theoretical and Computational Physics at Hong Kong, 
The University of Hong Kong, Hong Kong, China}
\author{Gang Chen}
\email{gangchen@hku.hk}
\affiliation{Department of Physics and HKU-UCAS Joint Institute 
for Theoretical and Computational Physics at Hong Kong, 
The University of Hong Kong, Hong Kong, China}  

\date{\today}

\begin{abstract}
Motivated by the discovery of the quantum anomalous Hall effect in 
Cr-doped \ce{(Bi,Sb)2Te3} thin films, we study the generic states for magnetic 
topological insulators and explore the physical properties for 
both magnetism and itinerant electrons. 
First-principles calculations are exploited to investigate 
the magnetic interactions between magnetic Co atoms 
adsorbed on the \ce{Bi2Se3} (111) surface. 
Due to the absence of inversion symmetry on the surface,
 there are Dzyaloshinskii-Moriya-like twisted spin interactions 
 between the local moments of Co ions. 
 These nonferromagnetic interactions twist the collinear 
 spin configuration of the ferromagnet and generate 
 various magnetic orders beyond a simple ferromagnet. 
 Among them, the spin spiral state generates alternating 
 counterpropagating modes across each period of spin states, 
 and the skyrmion lattice even supports a chiral mode 
 around the core of each skyrmion.
  The skyrmion lattice opens a gap at the surface Dirac point, 
  resulting in the anomalous Hall effect. 
  These results may inspire further experimental 
  investigation of magnetic topological insulators.
\end{abstract}

\maketitle

\section{introduction}

Topological insulators (TIs) have attracted tremendous research interest 
in the past decade, owing to their fundamental physics and potential 
applications for next-generation devices~\cite{Hasan2010,Qi2011}. 
Three-dimensional (3D) TIs have a gapped electron structure 
in the bulk and a metallic surface state whose existence 
is protected by the nontrivial topology of the bulk bands 
under time-reversal symmetry (TRS)~\cite{Fu2007,Moore2007,Qi2008}. 
The metallic surface state consists of an odd number of Dirac electrons. 
A unique nature that makes it different from an ordinary 
surface state is spin-momentum locking. 
The spin direction of the Dirac electron is dictated by its momentum direction, 
and vice versa. It cannot be gapped by nonmagnetic impurities because 
the TRS is still respected, and the bulk band topology remains nontrivial. 
In contrast, magnetism from the magnetic impurities would break the TRS. 
With a ferromagnetic order, the surface Dirac fermion
 acquires a mass, as demonstrated by experiments~\cite{Chen2010}. 
 Such gap opening by ferromagnetic impurities can lead to various interesting 
 physical phenomena, such as the half-integer quantum Hall effect~\cite{Qi2008}, the topological magneto-electric effect~\cite{Qi2008,Essin2009}, the induced magnetic monopole~\cite{Qi2009}, the quantum anomalous Hall effect~\cite{Haldane1988,YuRui2010,Chang2013}, and so on.

On the other hand, the TI surface itself breaks the spatial inversion symmetry 
between two magnetic adatoms. 
As a result, the magnetic interaction between the adatoms 
involves not only the Heisenberg interaction 
but also the Dzyaloshinskii-Moriya interaction (DMI)~\cite{Dzyaloshinsky1958,Moriya1960}. 
The latter is an effect rooted in spin-orbit coupling (SOC)
that cannot be ignored on the TI surface, 
and is nonvanishing in noncentrosymmetric systems in general. 
TIs such as \ce{Bi2Se3}~\cite{Zhang2009} have large SOC, 
so that their bulk bands are inverted, 
resulting in their nontrivial band topology. It 
is conceivable that such a large SOC can 
bring about significant DMI components in the adatom interactions. 
In fact, strong DMI has been found in the Co/Pt interface~\cite{Emori2014,Pizzini2014,Ryu2014,Belmeguenai2015,Yang2015}, 
where the heavy Pt atoms provide strong SOC to induce DMI between magnetic Co atoms. With the DMI, 
the magnetic adatoms on the TI surface could develop
much richer magnetic orders, such as the spin density wave, 
spin spirals, and the skyrmion lattice~\cite{Skyrme1962,Muhlbauer2009,Yu2010,Munzer2010,Heinze2011,Yu2011}, 
than a simple ferromagnet. 
The DMI between Fe adatoms on \ce{Bi2Se3} was 
calculated to be quite large~\cite{Li2012}. Experimental 
evidence has been found for the existence of magnetic skyrmions 
at the interface of ferromagnetic \ce{Cr2Te3} and TI \ce{Bi2Te3}~\cite{Chen2019}. Besides, a phase diagram that contains noncollinear 
magnetic structures 
such as spin spirals and the skyrmion lattice still needs to be established.

Conversely, the magnetic structures have an impact on the TI surface states. 
The low-energy spectrum may no longer be a massless Dirac cone, 
and the wavefunctions may not be uniformly distributed in space (on a large scale compared with the lattice constant) anymore. 
It was found that although out-of-plane ferromagnetically ordered surface impurities generate a band gap on the surface, the domain wall between up and down magnetization supports one-dimensional gapless chiral modes~\cite{Liu2009}. 
Magnetic textures such as domain walls and skyrmions become 
electrically charged when coupled to a 3D TI 
surface~\cite{Nomura2010,Hurst2015}. 
Magnetic skyrmions can give rise to bound states on TI surfaces~\cite{Andrikopoulos2016}. In regard to transport properties, 
the single magnetic skyrmion was found to induce an anomalous Hall 
effect on TI surfaces that is different from the conventional 
topological Hall effect~\cite{Araki2017}. The skew scattering 
from the skyrmion is robust against geometric deformation~\cite{Wang2020}. 
Most of these analyses focused on Bloch-type skyrmions,
and approximations such as hard-wall approximations were employed 
to model the skyrmion structure, inevitably losing the twisting 
information of the skyrmion.

In this paper we explore the phase diagram of a magnetic cobalt adatom 
lattice on a TI surface and investigate the effect of different twisted magnetic 
structures on the TI surface states. The remaining parts of the paper are 
organized as follows. 
In Sec.~\ref{sec: Geometric and magnetic properties of adatom},
we describe the results of first-principles calculations on the 
geometry of cobalt atoms adsorbed on the surface of a typical TI, 
\ce{Bi2Se3}. The magnetic properties of Co atoms are also discussed. 
In Sec.~\ref{sec: Magnetic interactions of adatoms and magnetic 
phase diagram of adatom lattice}, we calculate the magnetic interactions 
between the Co adatoms. Then Ginzburg-Landau theory is used to 
analyze the magnetic phase diagram of a Co adatom lattice. 
Magnetic structures of spin spirals and skyrmions are found 
to be achievable. In Sec.~\ref{sec: Dirac electrons couple to adatoms}, 
we numerically solve the problem of TI surface Dirac electrons 
coupled to the twisted magnetic structures. 
The electronic states under the magnetic background of a spin spiral, 
a single skyrmion, and a skyrmion lattice are discussed. 
The paper is concluded in Sec.~\ref{sec: Discussions and conclusions}.


\section{Geometric and magnetic properties of the adatom}
\label{sec: Geometric and magnetic properties of adatom}

\begin{figure}[t]
    \includegraphics[width=7.0cm]{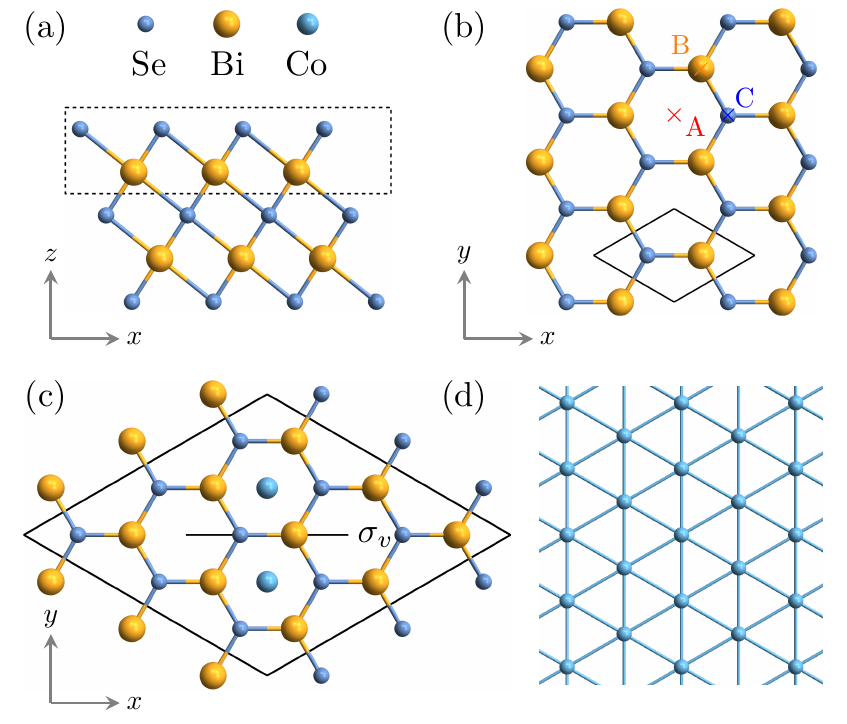}
    \caption{(a) A side view of a QL of \ce{Bi2Se3}. (b) Viewing from the $ z $ direction, the two topmost atomic layers enclosed by a dashed rectangle in (a) form a hexagonal lattice with the primitive cell indicated by a rhombus. Three typical adsorption sites are marked by colored crosses, in which site A is most stable. (c) The $ 3 \times 3 $ supercell used for calculating the magnetic interactions between two Co adatoms connected by a mirror plane $\sigma_v$. (d) A triangular lattice formed by the Co adatoms.}
    \label{fig: lattice_geometry}
\end{figure}

The material \ce{Bi2Se3} has a layered structure consisting of Se-Bi-Se-Bi-Se 
quintuple layers (QLs) stacking along the crystallographic $ c $ axis. 
Each atomic layer forms a triangular lattice, and they stack in the 
ABCAB sequence within each QL [see Fig.~\ref{fig: lattice_geometry}(a)]. 
The two topmost atomic layers of Bi and Se, enclosed by the dashed 
rectangle in Fig.~\ref{fig: lattice_geometry}(a), make up a honeycomb lattice 
when viewed from the $z$ direction, as shown in Fig.~\ref{fig: lattice_geometry}(b). 
The equilibrium position of a magnetic cobalt atom adsorbed on the surface 
is determined by first-principles calculations (see details in Appendix~\ref{app: DFT_details}). 
It is found that among the three typical adsorption sites marked by colored crosses in Fig.~\ref{fig: lattice_geometry}(b), site A is the most stable one. The height of the Co atom is $ 0.24~\mathring{\text{A}} $ lower than the top Se layer.
Thus the Co adatom can be seen as an interstitial impurity occupying the hollow site buried a little bit into the \ce{Bi2Se3} (111) surface.

\begin{figure*}
	\includegraphics[width=17cm]{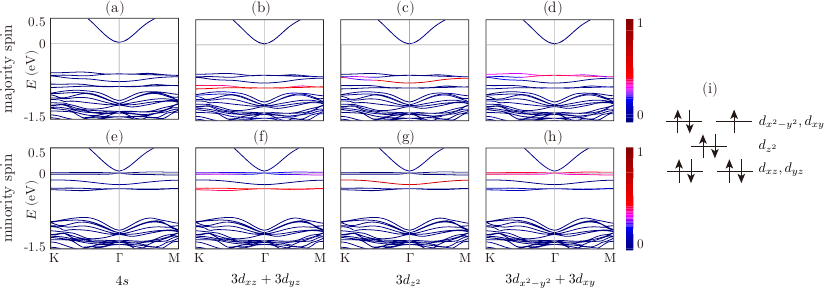}
	\caption{The band structures of the (a)--(d) majority spin and (e)--(h) minority spin of a Co atom adsorbed on a $ 3 \times 3 $ supercell of \ce{Bi2Se3} without the spin-orbit interaction. The color represents the weight of different Co atomic orbitals: $ 4s $ [(a) and (e)], $ 3d_{xy}+3d_{yz} $ [(b) and (f)], $ 3d_{z^2} $ [(c) and (g)], and $ 3d_{x^2-y^2}+3d_{xy} $ [(d) and (h)]. The Fermi level is shifted to $ E=0 $. (i) schematically represents the occupancy of the Co $ 3d $ shell.}
	\label{fig: Hund}
\end{figure*}

In the absence of spin-orbit coupling (SOC), the magnetic moment of the Co adatom is $1.0 \ \mu_B$ from our first-principles calculation. 
To investigate the electron configuration, the band structure of a $3 \times 3$ supercell is plotted, where the weight of different Co atomic orbitals is color coded as presented in Figs.~\ref{fig: Hund}(a)--\ref{fig: Hund}(h). 
It can be seen that the $3d$ states of the Co atom lie 
inside the bulk gap of \ce{Bi2Se3}.
Due to the $C_{3v}$ point-group symmetry of the adsorption site $A$, the five $3d$ orbitals split into three sets in their energies: $ \{d_{xz}, d_{yz}\} $, $ \{d_{z^2}\} $, and $ \{d_{x^2-y^2}, d_{xy}\} $.
In contrast to the empty $4s$ orbitals, most of the $3d$ orbitals are fully occupied except the $ \{d_{x^2-y^2}, d_{xy}\} $ manifold, which has an unpaired electron. 
From this observation, it can be inferred that the electron configuration of the Co adatom is $ 4s^{0}3d^{9} $, and the filling is schematically shown in Fig.~\ref{fig: Hund}(i).
There is no doping from Co to \ce{Bi2Se3} when the SOC in the system is shut off, only that the two $ 4s $ electrons are transferred into the $ 3d $ shell. 
This is different from the isolated Co atom, which has the electron configuration $ 4s^{2}3d^{7} $. 
The existence of only one unpaired electron explains the $1.0\ \mu_B$ magnetic moment.

\begin{figure}
	\includegraphics[width=8.2cm]{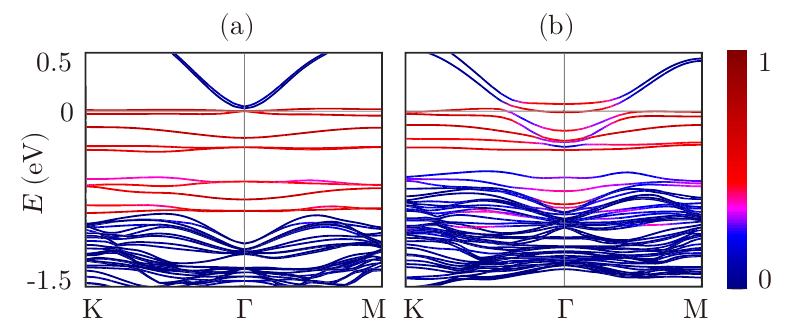}
	\caption{Comparison of the band structure of a Co atom adsorbed on a $ 3 \times 3 $ supercell of  \ce{Bi2Se3} (a)  without and (b) with spin-orbit interaction. The weight of the Co atomic orbitals is color-coded. The Fermi level is shifted to $ E=0 $.}
	\label{fig: bands}
\end{figure}

The SOC slightly raises the magnetic moment of the Co adatom to about $1.2\ \mu_B$.
The increase can be attributed to the doping effect of the Co electrons into the surface state of \ce{Bi2Se3}. 
To see this, the band structures without and with SOC are compared in Fig.~\ref{fig: bands}.
The weight of the Co atomic orbitals is color coded. 
Without SOC, the bands originating from the Co atom are nearly flat, indicating that the Co $ 3d $ electrons are well isolated. 
When SOC is turned on, the Co bands overlap and hybridize with those from \ce{Bi2Se3}.
The band hybridization leads to a fraction of the Co electrons doping into the \ce{Bi2Se3} surface.
Because of such doping, the unpaired part of the Co electrons becomes larger considering Hund's rule, which accounts for the larger magnetic moment.

In Fig.~\ref{fig: bands} one can further see that the surface state of \ce{Bi2Se3} is gapped, and only the upper massive Dirac cone is in the bulk band gap. This is because we use a slab of one quintuple layer in our first-principles calculations. The surface states of the top and the bottom layer overlap in the bulk and hybridize, leading to the gap opening~\cite{Zhang2010,Sakamoto2010}.

\section{Magnetic interactions of adatoms and magnetic phase diagram of the adatom lattice}\label{sec: Magnetic interactions of adatoms and magnetic phase diagram of adatom lattice}

To calculate the magnetic interactions between two Co adatoms 
on neighboring A sites, we use a $ 3 \times 3 $ supercell as shown 
in Fig.~\ref{fig: lattice_geometry}(c). 
The spin-spin interactions between them can be described 
by the Hamiltonian
\begin{equation}\label{eq: spin-spin interactions}
    H_{12}=\vec{S}_{1}\cdot\tensor{J}\cdot\vec{S}_{2}.
\end{equation}
The system has a mirror symmetry $ \sigma_{v} $ 
with respect to the plane which perpendicularly 
bisects the line joining atoms 1 and 2. 
With such a symmetry constraint, 
the tensor $ \tensor{J} $ takes the general form
\begin{equation}\label{eq: interaction_matrix}
    \tensor J=
    \begin{bmatrix}
    J_{xx} & D_{z} & \Gamma_{xz} \\
    -D_{z} & J_{yy} & D_{x} \\
    \Gamma_{xz} & -D_{x} & J_{zz}
    \end{bmatrix},
\end{equation}
in which $ J_{ii} $ ($ i=x,y,z $) are Heisenberg interactions, $ D_x $ and $ D_z $ are DMIs, and $ \Gamma_{xz} $ is the off-diagonal pseudodipolar interaction.

The results of the parameters in $ \tensor{J} $ are listed in Table~\ref{tab: Jij} (see methods in Appendix~\ref{app: spin_interactions}). 
For convenience, we have rescaled the interaction parameters for a normalized spin $ |\vec{S}| = 1 $. It can be seen that the symmetry requirement of $ \Gamma_{xy}=\Gamma_{yz}=D_{y}=0 $ is confirmed by first-principles calculations. 
It is clear that the Co adatoms are mainly coupled by the Heisenberg ferromagnetic interaction. 
Besides, there are the DMI and pseudodipolar interaction. Both of them are relativistic SOC effects.
The former is made nonzero by the broken inversion symmetry of the TI surface. 
Among them, the DMI can lead to more interesting magnetic structures than ferromagnetism.
In the following, we assume a high coverage of Co adatoms located on the A sites and arranged in a triangular lattice as shown in Fig.~\ref{fig: lattice_geometry}(d).

\begin{table}
    \begin{ruledtabular}
        \begin{tabular}{ccccccccc}
            $ J_{xx} $ & $ J_{yy} $ & $ J_{zz} $ & $ \Gamma_{xy} $ & $ \Gamma_{xz} $ & $ \Gamma_{yz} $ & $ D_{x} $ & $ D_{y} $ & $ D_{z} $ \\
            $-56.12$ & $-57.19$ & $-60.12$ & $0.00$ & $-0.38$ & $0.00$ & $-0.41$ & $0.00$ & $1.03$
        \end{tabular}
    \end{ruledtabular}
\caption{First-principles calculation results of the magnetic interaction parameters in Eq.~\eqref{eq: spin-spin interactions}, in units of meV. The interacting spins have been normalized to unit vectors, $ |\vec{S}| = 1 $.}
\label{tab: Jij}
\end{table}

To discuss the magnetic ground state, we derive the continuum model from the microscopic magnetic interaction model and make use of the Ginzburg-Landau theory. 
In the long wavelength limit, one can expand the spin distribution to the second order,
\begin{equation}
    S_{\mu}(\vec{r}+\delta\vec{r})\approx S_{\mu}(\vec{r})+\delta\vec{r}\cdot\vec{\nabla}S_{\mu}(\vec{r})+\frac{1}{2}(\delta\vec{r}\cdot\vec{\nabla})^{2}S_{\mu}(\vec{r}),
\end{equation}
and obtain the zero-temperature free energy under external magnetic field $ \vec{B} = (0,0,B) $ up to quadratic order of $\vec{S}(\vec{r})$
\begin{align}\label{eq: continuum model}
    F & \simeq \int d^{2}\vec{r}\Bigg\{\frac{\tilde{J}_{1}}{2}\left[(\partial_{x}S_{x})^{2}
    +(\partial_{y}S_{y})^{2}\right]+\frac{\tilde{J}_{2}}{2}\left[(\partial_{x}S_{y})^{2}
    +(\partial_{y}S_{x})^{2}\right]
    \notag \\
    & +\frac{\tilde{J}_{3}}{2}
    \left[(\partial_{x}S_{z})^{2}+(\partial_{y}S_{z})^{2}\right]+(\tilde{J}_{1}-\tilde{J}_{2})\partial_{x}S_{x}\partial_{y}S_{y}
    \notag \\
    & +\tilde{\Gamma}(\partial_{x}S_{x}\partial_{x}S_{z}
    -2\partial_{x}S_{y}\partial_{y}S_{z}-\partial_{y}S_{x}\partial_{y}S_{z}) \notag \\
    & +2\tilde{D}(S_{x}\partial_{x}S_{z}+S_{y}\partial_{y}S_{z})
    +\tilde{A}S_{z}^{2}+\tilde{C}S^{2}-BS_{z} \Bigg\},
\end{align}
where the coefficients are related to the microscopic parameters as
\begin{align}
    \tilde{J}_{1} & =-\frac{\sqrt{3}}{2}\left(J_{xx}+3J_{yy}\right),\quad\tilde{J}_{2}=-\frac{\sqrt{3}}{2}\left(3J_{xx}+J_{yy}\right), \label{eq: micro-macro relations-1}\\
    \tilde{J}_{3} & =-2\sqrt{3}J_{zz},\quad\tilde{\Gamma}=\sqrt{3}\Gamma_{xz},\quad\tilde{D}=\frac{2\sqrt{3}}{a}D_{x}, \\
    \tilde{A} & =\frac{2\sqrt{3}}{a^{2}}\left(2J_{zz}-J_{xx}-J_{yy}+\frac{2}{3}A\right),\\
    \thinspace\tilde{C} & =\frac{2\sqrt{3}}{a^{2}}\left(J_{xx}+J_{yy}\right). 
\label{eq: micro-macro relations-2}
\end{align}
Here, $ a = 4.11\ \mathring{\text{A}} $ is the adatom lattice constant, and a contribution from single-ion anisotropy $A$ has been incorporated in $\tilde{A}$. The first-principles calculation finds a tiny $ A = -0.04\ \text{meV} $.
Note that the low-energy continuum model, Eq.~\eqref{eq: continuum model}, has the same form as in Ref.~[\onlinecite{Li2021}] except for the lattice-dependent coefficients in Eqs.~\eqref{eq: micro-macro relations-1}--\eqref{eq: micro-macro relations-2}.
This is because in the Ginzburg-Landau theory  
the underlying microscopic differences 
are coarse grained and it relies solely on the $C_{3v}$ point-group symmetry.

The magnetic structures of the free energy with the form of Eq.~\eqref{eq: continuum model} have been discussed in detail~\cite{Li2021}. 
Here, we review the main conclusions. 
The free energy can be split into two parts, $ F = F_0 + F' $, in which $ F_0 $ contains the $ \tilde{J}_{1} $, $ \tilde{J}_{2} $, $ \tilde{J}_{3} $, $ \tilde{D} $, and $ B $ terms.
One can make the approximation $ \tilde{J}_{1}\approx\tilde{J}_{2}\approx\tilde{J}_{3} $ and treat the difference $ F' $ as a perturbation according to their values in Table~\ref{tab: Jij}. 
The ground state of $ F_0 $ is a spin spiral
\begin{equation}\label{eq: spin spiral}
\vec S(\vec r)=\left[\phi_{1} \sin(\vec q\cdot\vec r)\cos\theta,\phi_{1}\sin(\vec q\cdot \vec r)\sin\theta,\phi_{1}\cos(\vec q\cdot\vec r)+\phi_{0}\right]
\end{equation}
with the propagation vector $ \vec q=Q\left(\cos\theta,\sin\theta,0\right) $, $ Q=|\tilde{D}/\tilde{J}| $. 
The additional ferromagnetic component $ \phi_{0} $ comes from the partial spin polarization under the external magnetic field and satisfies the global normalization constraint~\cite{Park2011}
\begin{equation}
\langle|\vec S\left(\vec r\right)|^{2}\rangle =\phi_{1}^{2}+\phi_{0}^{2}=1.
\end{equation}
The contour degeneracy of arbitrary $ \theta $ 
for the free energy is quite similar to the case 
in spiral spin liquids~\cite{Yao2021}. 
The inclusion of $ F' $ leads to a hexagonal warping 
in the free energy landscape and six discrete 
propagation directions favored by the ground-state spin spiral.

As more terms beyond quadratic order are considered, spin spirals with different directions of propagation vectors can interact with each other, resulting in the superposition of several spin spirals. 
The leading term of such a correction is a quartic one 
${ \Delta F\propto\int d^{2}\vec r\left|\vec S\left(\vec r\right)\right|^{4} }$. 
It makes possible the ground state being the skyrmion lattice (SkX)
\begin{equation}\label{eq: SkX}
\vec S\left(\vec r\right)=\sum_{i=1}^{6}\phi_{i}\me^{\mi\vec q\cdot\vec r}\frac{\vec e_{i}}{\sqrt{2}}+\phi_{0}\vec e_{z},
\end{equation}
where ${ \vec q_{i}=Q(\cos\theta_{i},\sin\theta_{i},0) }$ and 
$ \vec e_{i}= \frac{1}{\sqrt{2}}(-\mi\cos\theta_{i},-\mi\sin\theta_{i},1) $. 
The $ \theta_{i} $'s and $ \phi_{i} $'s have the relationship $ \theta_{2}=\theta_{1}+2\pi/3 $, $ \theta_{3}=\theta_{1}+4\pi/3 $, $ \theta_{i+3}=\theta_{i}+\pi $, and $ \phi_{i+3}=\phi_{i}^{*} $ ($ i=1,2,3 $). 
The SkX is composed of three spin spirals whose propagation vectors have a relative angle of $ 120^\circ $ with one another. 
Their amplitudes are subject to the constraint $ \sum_{i=1}^{3}|\phi_{i}|^{2}+\phi_{0}^{2}=1 $.

\begin{figure}
	\includegraphics[width=8.6cm]{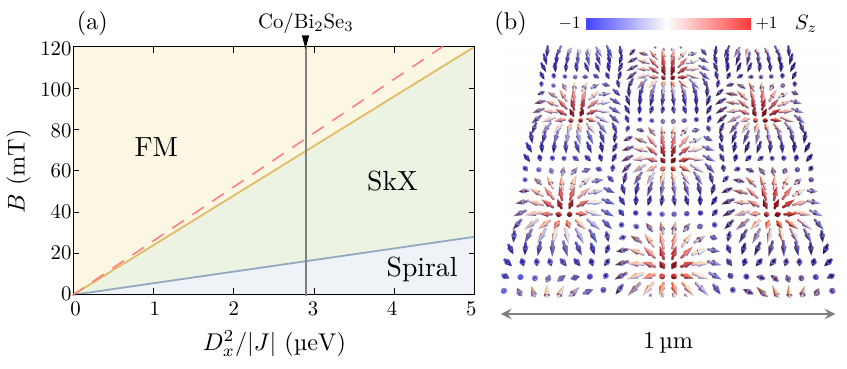}
	\caption{(a) The phase diagram for the spin spiral, skyrmion lattice (SkX), and ferromagnetic (FM) structure. The small yellow region below the red dashed line is a magnetic structure in which the spin directions are twisted but can be continuously connected to the FM phase. The calculated material parameters are shown by the vertical line. (b) A typical spin configuration of the SkX phase, where the out-of-plane component $S_z$ is indicated by colors.}
	\label{fig: phase_diagram_SkX}
\end{figure}

After comparing the free energy including the quartic term between the spin spiral, Eq.~\eqref{eq: spin spiral}, and the SkX, Eq.~\eqref{eq: SkX}, it was found that when the external magnetic field strength lies in the range
\begin{equation}\label{eq: SkX range}
\frac{5\tilde{D}^{2}/\tilde{J}}{\sqrt{457}}<B<\tilde{D}^{2}/\tilde{J},
\end{equation}
the SkX will have lower energy than the spin spiral~\cite{Li2021}. 
Here, $ \tilde{J} $ is the average of $ \tilde{J}_1 $, $ \tilde{J}_2 $, and $ \tilde{J}_3 $. 
We draw the phase diagram on the $ B $--$ D_{x}^{2}/|J| $ plane in \Fig\ref{fig: phase_diagram_SkX}(a), in which $ B $ is converted to units of milliteslas and $ J $ is the average of $ J_1 $, $ J_2 $, and $ J_3 $. 
The calculated material parameters are indicated by the vertical black line. 
The magnetic field strength corresponding to the SkX phase is about tens of milliteslas, which is quite accessible in the laboratory.
The wavelength of the spin spiral is $ \lambda=2\pi/Q=364~\text{nm} $. 
A typical spin distribution of the SkX phase is presented in \Fig\ref{fig: phase_diagram_SkX}(b). 
The distance between two adjacent skyrmion centers is $ 2\lambda/\sqrt{3}=421~\text{nm} $. 
The spin spiral and the SkX are both of the N\'eel type, with the spins rotating in the plane spanned by the wave vector and the surface normal. 
Note that although the external magnetic field influences the energetics among different magnetic structures and adds some ferromagnetic spin components into the spin spiral and SkX, it does not alter their periods.

\section{Dirac electrons couple to magnetic textures}
\label{sec: Dirac electrons couple to adatoms}

In the previous section, we have investigated the twisting effect 
of Dirac electrons on the magnetic interactions between adsorbed 
Co atoms. 
It gives rise to versatile magnetic orders as shown in 
Fig.~\ref{fig: phase_diagram_SkX}(a).
Once the spin spiral or SkX is formed on the adatom lattice, 
the Dirac electrons could interact with the large-scale magnetic 
texture $\vec{S}(\vec{r})$  
conversely via the Kondo (or Hund's) coupling.
In general, this coupling tends to align the spin of the electrons 
with the orientation
of local moment, an inverse process of the spin transfer torque effect. 
However, further complexities would inevitably emerge when the spin-momentum locking is incorporated for the Dirac electrons.
Such a process can be modeled by the following Hamiltonian:
\begin{equation}\label{eq: electron-spin interaction}
H=v_{\text{F}}\left(\vec{p}\times\vec{\sigma}\right)\cdot\hat{\vec z}-J_{\text{ex}}\vec S\left(\vec r\right)\cdot\frac{\vec{\sigma}}{2}.
\end{equation}
The first term is a Rashba-type Hamiltonian 
describing the surface state of \ce{Bi2Se3}~\cite{Hasan2010}, 
and the second term is Kondo (or Hund’s) coupling 
between the electron spin and the spin of the magnetic adatom. 
Because the ferromagnetic exchange coupling $ J_{\text{ex}} $ is typically much larger than the Zeeman energy arising from the external magnetic field, we neglect the latter hereinafter for the sake of simplicity. 
In this section, we explore the fate of Dirac electrons under the fixed back ground spin texture $ \vec{S}(\vec{r}) $ of a spin spiral, a single skyrmion, and a skyrmion lattice.

\subsection{The spin spiral case}

\begin{figure}
	\includegraphics[width=8.2cm]{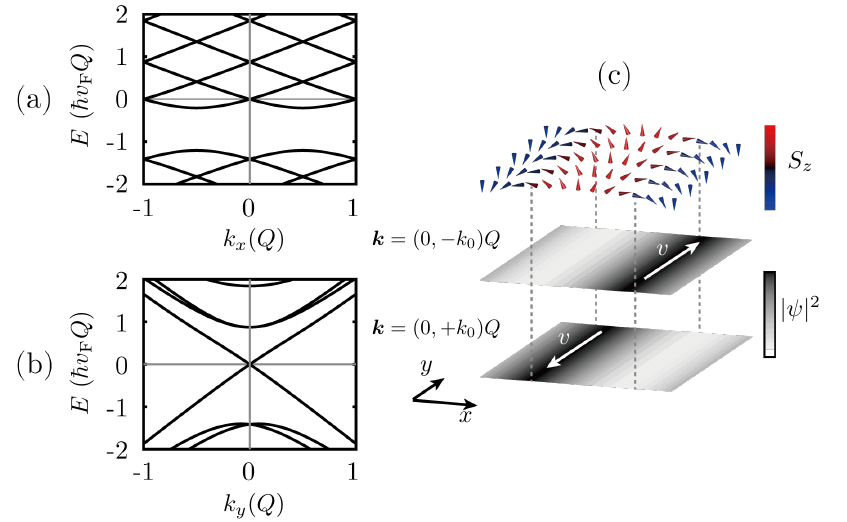}
	\caption{The band structures of Dirac electrons on the TI surface coupled to a spiral magnetic structure, with (a) $ k_y = 0 $ and (b) $ k_x = 0 $. (c) The spin configuration of the spiral magnetic structure is shown in the upper panel. The modulus squared of the wave functions of the valence states at $ \vec{k}=(0,-k_0)Q $ and $ \vec{k}=(0,+k_0)Q $ (with $ k_0 =0.1 $) are shown in the middle and lower panels, respectively. The directions of the group velocity are represented by the white arrows.}
	\label{fig: spiral_surface_state}
\end{figure}

We first consider a spin spiral propagating along the $ x $ direction, 
whose spin configuration is
\begin{equation}
\vec{S}\left(\vec r\right)=\left(\sin Qx,0,\cos Qx\right).
\end{equation}
Because of the periodicity along the $ x $ and $ y $ directions, 
the electron wave function $ \psi(\vec{r}) $ can be expanded 
by plane waves with a cutoff $N$,
\begin{equation}
    \psi\left(\vec r\right)=\me^{\mi\left(k_{x}x+k_{y}y\right)}\sum_{|n|\le N}\me^{\mi nQx}
    \begin{bmatrix}
        u_n \\
        v_n
    \end{bmatrix}.
\end{equation}
By taking the coupling parameter $ {J_{\text{ex}}= 1} $ 
in units of  $ \hbar v_{\text{F}} Q $, we solve this eigenvalue problem in $N=5$.
The Fermi energy is determined by requiring half filling of the energy bands, 
and is tuned to $ E=0 $. 
Then without the coupling to the spin spiral, 
the Fermi energy lies at the Dirac point of the surface state.

From the band structures shown in Figs.~\ref{fig: spiral_surface_state}(a) and \ref{fig: spiral_surface_state}(b), one can see that after coupling to the spin spiral,
 the TI surface remains a semimetal, with the Fermi energy still lying at a Dirac cone. 
The spatial distributions of two wave functions with $ {\vec{k} = (0,\pm k_0)Q }$ 
on the negative branch of the Dirac cone are presented in 
\Fig~\ref{fig: spiral_surface_state}(c). Here, ${ k_0 = 0.1 }$. 
By comparing with the spin configuration of the spiral structure, 
it can be seen that the wave functions localize around the domain walls 
with ${ S_z = 0 }$. 
The wave function at the domain wall with ${ S_x > }0 $ (${ S_x < 0 }$) 
has a velocity ${ v_y > 0 }$ (${ v_y < 0} $). 
The spatial distribution and velocity of the states near 
the Fermi energy are reminiscent of the chiral edge states 
of the quantum Hall effect. For a small $ Q $, the magnetic 
structure varies rather smoothly, so that an electron may 
only feel the environment surrounding it. 
In the $ {S_z \ne 0} $ areas, the electron behaves 
as if its band structure has a Dirac mass whose sign is 
determined by the sign of $ S_z $. 
In the transition region where $ S_z $ changes its sign, 
the absolute value of Dirac mass decreases to zero and then increases. 
The existence of gapless states at the domain walls can 
be understood qualitatively in this way.

We further comment on the possibility of thermal fluctuations 
in the spin spiral state. As the energy scale of the exchange is 
much lower than the electron bandwidth, the thermal fluctuations 
would affect the magnetism first. It is known that, 
the thermal fluctuations often convert the 
coplanar spin spiral state into a collinear spin density wave state 
before the system enters the paramagnetic state. If the collinear 
spin density wave state modulates with the $S_z$ component,
the chiral edge mode associated with the domain walls
persists until the system becomes paramagnetic.

It is illuminating to make a connection with the early work on 
magnetic domains on the surface of topological insulators. 
There, on the domain wall between two neighboring
ferromagnetic domains, there exists a conducting chiral edge mode. 
This is because the neighboring ferromagnetic domains 
with opposite magnetic orders perpendicular to the 
surface have opposite Chern numbers~\cite{Hasan2010}.  
This domain chiral edge mode has been
realized experimentally
on the surface of magnetic topological insulators~\cite{2017Sci...358.1311Y,Checkelsky2012},
and with
a more controlled method with the strong permanent magnet placed atop the 
GaAs/AlGaAs quantum Hall heterojunction~\cite{2019NatCo..10.4565J}. 
The spin modulation of the spin spiral state with one spiral period can 
be approximately viewed as one domain wall, and there will naturally be 
a chiral edge mode associated with it.

\subsection{The single-skyrmion case}

\begin{figure}[t]
    \includegraphics[width=8.5cm]{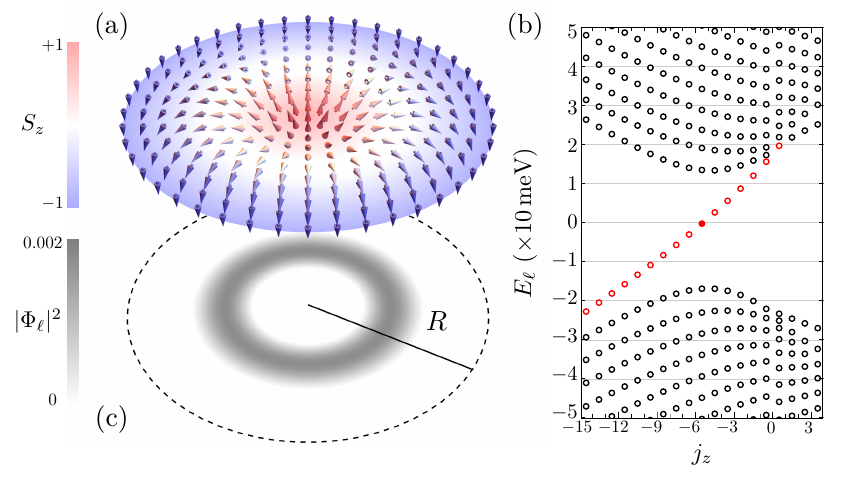}
    \caption{A N\'{e}el-type skyrmion on a disk of radius $R$. The out-of-plane components $S_z$ are indicated by colors. (b) The low-energy spectrum of the surface Dirac electrons coupled to a single skyrmion, showing chiral edge states (colored in red) inside the gap. (c) The probability density $|\Phi_{\ell}|^2$ of the edge state with near-zero energy, marked as a solid red circle in (b).}
    \label{fig: SingleSkyrmion}
\end{figure}

Being an example of a topologically nontrivial magnetic texture, 
a skyrmion could influence the flow of electron spins and even 
nucleate itinerant electrons through the spin transfer torque effect. 
This is quite different from the spin spiral case. 
Before considering the SkX, we first clarify the physics of 
Dirac electrons within a single skyrmion in order to get some insight. 
To achieve this, the effective Hamiltonian in Eq.~\eqref{eq: electron-spin interaction} is applied to an open system with the disk geometry of radius $R$ as shown in Fig.~\ref{fig: SingleSkyrmion}(a).
Note that only the spin momentum $\vec{\sigma}$ 
is involved in the exchange interaction and the possible 
contribution from the orbital momentum ${\vec{L} = \vec{r} \times \vec{p}}$ 
due to the nucleation is not considered at this stage.
As analyzed in the previous section, the skyrmion 
on the surface of Co/\ce{Bi2Se3} is of the N\'{e}el type 
and hence can generally be written in the polar 
coordinates $\vec{r} = (r, \theta)$ as
\begin{equation}
	\vec{S}(\vec{r})=\left[\cos\theta \sin\left(\frac{\pi}{R}r\right), \sin\theta \sin\left(\frac{\pi}{R}r\right), \cos\left(\frac{\pi}{R}r\right)\right],
\end{equation}
so that at the skyrmion core ${r=0}$ we have $\vec{S}(\vec{r})=[0,0,1]$
and at the skyrmion boundary ${r= R}$ we have $\vec{S}(\vec{r})=[0,0,-1]$. 

It can be verified that the $z$ component of the total angular momentum 
$J_z=-\mi\hbar\partial_{\theta} + \hbar/2 \sigma_z$ commutes 
with the effective Hamiltonian in Eq.~\eqref{eq: electron-spin interaction} and 
thus provides a good quantum number $j_z$.
The wave function has the general form
\begin{equation}
    \Phi_{\ell}(r,\theta) = 
    \begin{bmatrix}
        u_{\ell} (r) e^{\mi \ell \theta} \\
        v_{\ell} (r) e^{\mi (\ell + 1) \theta}
    \end{bmatrix}
\end{equation}
with an integer $\ell$, satisfying  
${J_z\Phi_{\ell}(r,\theta) = j_z \hbar \Phi_{\ell}(r,\theta)}$ 
with ${j_z = \ell + 1/2}$. The Hamiltonian 
in Eq.~\eqref{eq: electron-spin interaction} can be 
further separated into angular and radial parts, 
and the latter reads
\begin{equation}
    H_{\ell} 
    \begin{bmatrix}
        u_{\ell}(r) \\
        v_{\ell}(r)
    \end{bmatrix}
    = E_{\ell}
    \begin{bmatrix}
        u_{\ell}(r) \\
        v_{\ell}(r)
    \end{bmatrix},
\end{equation}
where
\begin{widetext}
\begin{equation}
    H_{\ell} = 
	\begin{bmatrix}
        -\frac{J_{\text{ex}}}{2}\cos\left(\frac{\pi}{R}r\right) & -\hbar v_{\text{F}}\left(\frac{\partial}{\partial r}+\frac{\ell+1}{r}\right)-\frac{J_{\text{ex}}}{2}\sin\left(\frac{\pi}{R}r\right) \\
       \hbar v_{\text{F}}\left(\frac{\partial}{\partial r}-\frac{\ell}{r}\right)-\frac{J_{\text{ex}}}{2}\sin\left(\frac{\pi}{R}r\right) & +\frac{J_{\text{ex}}}{2}\cos\left(\frac{\pi}{R}r\right)
    \end{bmatrix}.
\end{equation}
\end{widetext}
The radial Hamiltonian can be solved in each subspace 
of fixed angular momentum quantum number $\ell$ by 
expanding the radial wave functions $u_{\ell}(r)$ and 
$v_{\ell}(r)$ in the Fourier-Bessel series with a cutoff 
$N$ \cite{PhysRevX.9.011033},
\begin{equation}
    u_{\ell}(r) = \sum_{n = 1}^{N} u_{n} \varphi_{\ell,n}(r), \quad  v_{\ell}(r) = \sum_{n = 1}^{N} v_{n} \varphi_{\ell + 1, n}(r).
\end{equation}
The orthonormal basis is defined by 
\begin{equation}
    \varphi_{\ell, n}(r) = \frac{\sqrt{2}}{R J_{\ell + 1}(j_{\ell,n})}J_{\ell}\left(j_{\ell,n}\frac{r}{R}\right), \quad n = 1, \ldots, N,
\end{equation}
where the parameter $j_{\ell,n}$ is the $n$th zero of the $\ell$-order Bessel function of the first kind $J_{\ell}(x)$.
The Dirichlet boundary condition has been assumed implicitly.
This approximation reduces the radial Hamiltonian to a $2N \times 2N$ matrix eigenvalue problem
\begin{equation}
    \begin{bmatrix}
       - C_{\ell,\ell} & T_{\ell,\ell + 1} - S_{\ell,\ell + 1} \\
       T_{\ell + 1,\ell} - S_{\ell + 1, \ell}  & + C_{\ell+1, \ell+1}
    \end{bmatrix}
    \psi_{\ell} = E_{\ell} \psi_{\ell},
\end{equation}
where $ \psi_{\ell}^{T} = (u_1, \ldots, u_N, v_1, \ldots, v_N) $.
The matrix elements of the Hamiltonian are given by 
\begin{align}
    (C_{\ell,\ell'})_{n,n'}  & = \frac{J_{\text{ex}}}{2} \int_{0}^{R} \cos\left(\frac{\pi}{R}r\right) \varphi_{\ell, n}(r) \varphi_{\ell', n'}(r) r dr, \\
    (S_{\ell,\ell'})_{n,n'} & = \frac{J_{\text{ex}}}{2} \int_{0}^{R} \sin\left(\frac{\pi}{R}r\right) \varphi_{\ell, n}(r) \varphi_{\ell', n'}(r) r dr, \\
    (T_{\ell, \ell + 1})_{n, n'} & = + \frac{2 \hbar v_{\text{F}}}{R} \frac{j_{\ell,n} j_{\ell + 1, n'}}{j_{\ell, n}^2 - j_{\ell + 1, n'}^2}, \\
    (T_{\ell + 1, \ell})_{n, n'} & = - \frac{2 \hbar v_{\text{F}}}{R} \frac{j_{\ell + 1, n} j_{\ell, n'}}{j_{\ell + 1, n}^2 - j_{\ell, n'}^2}.
\end{align}

The Fermi velocity of a \ce{Bi2Se3} surface Dirac electron is 
about ${\hbar v_{\text{F}} = 329\,\text{meV nm}}$. 
For numerical convenience, a characteristic length $r_0$
is defined such that the energy unit ${\hbar v_{\text{F}} / r_0 = 100\,\text{ meV}}$, 
which gives ${r_0 \approx 3.3\text{nm}}$. 
Based on the analysis of density functional theory (DFT) and Ginzburg-Landau theory, 
the typical radius of a single skyrmion is estimated to be ${R=60}$ in units of $r_0$. 
With the coupling ${J_{\text{ex}} = 40\,\text{meV}}$ and a cutoff ${N=100}$ 
on the Fourier-Bessel basis, the low-energy spectrum is computed 
and depicted in Fig.~\ref{fig: SingleSkyrmion}(b).
In the total angular momentum space, the Dirac electrons 
open an energy gap immediately, and a chiral edge model emerges 
concurrently inside. This can be understood through the skyrmion topology. 
The swirling structure of a skyrmion is characterized by the topological charge, 
which counts the times that $\vec{S}(\vec{r})$ wraps a virtual sphere.
For a single N\'{e}el-type skyrmion with a unit topological number 
shown in Fig.~\ref{fig: SingleSkyrmion}(a), the winding of spins divides 
the disk into two regions bounded by ${r=R/2}$. 
Both the net magnetization and its resulting Dirac mass 
have opposite signs inside and outside the boundary.
Therefore there must be a closing and reopening of the bulk energy gap 
when crossing from one region to another, 
which produces edge modes near the boundary.
In Fig.~\ref{fig: SingleSkyrmion}(c), the probability 
density $|\Phi_{\ell}|^2$ for one of such edge states 
with near-zero energy is shown and compared with 
the skyrmion spin texture.
In addition, the degeneracy of states with $\pm j_z$ 
is lifted by the coupling due to the spin-momentum locking, 
and only the one with smaller energy can be localized on the boundary.
This endows the edge mode with a chirality.

\subsection{The skyrmion lattice case}

To investigate the effect arising from the pure spin texture of the skyrmion lattice, we neglect the possible ferromagnetic component in the spin configuration. 
Although the SkX phase has a lower energy than the ferromagnetic and spin spiral phase only in the application of the external magnetic field, there are energy barriers among them, meaning that even though the magnetic field is turned off, the SkX state can still be metastable, with no ferromagnetic component.

\begin{figure}[t]
	\includegraphics[width=8.2cm]{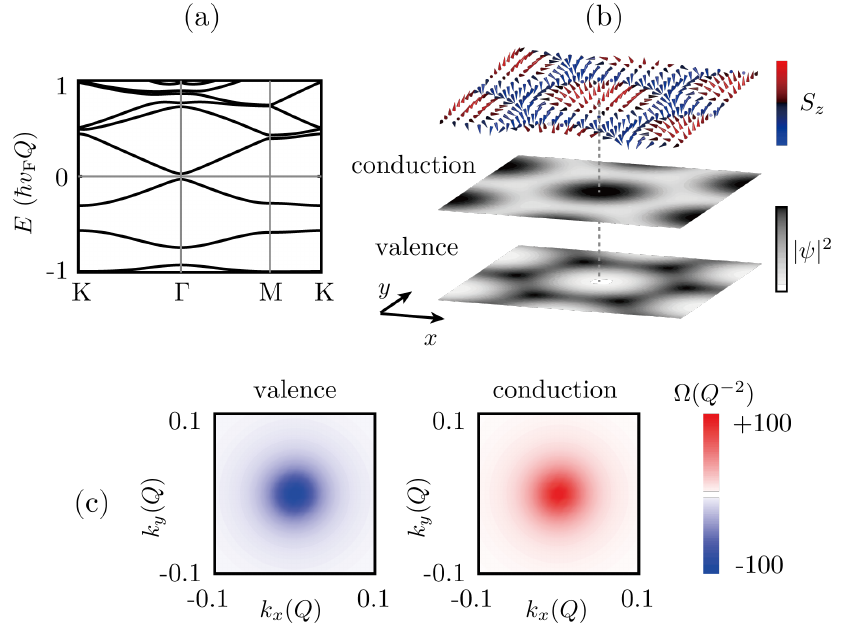}
	\caption{(a) The band structure of the topological surface state coupled to a skyrmion lattice. (b) The spin configuration of the skyrmion lattice is shown in the upper panel. The wave-function distributions of the conduction and valence states at the $ \Gamma $ point are shown in the middle and the lower panels, respectively. (c) The Berry curvature distributions of the valence and conduction bands near the Dirac gap are shown in the left and the right panels, respectively.}
	\label{fig: SkX_surface_state}
\end{figure}

The SkX solution of the Ginzburg-Landau free energy is given by Eq.~\eqref{eq: SkX}. After turning off the ferromagnetic component $ \phi_0 $, the spin distribution of the SkX under investigation is
\begin{multline}
\vec S\left(\vec r\right)= \Bigg[\frac{1}{\sqrt{3}}\left(\sin Qx+\sin\frac{Qx}{2}\cos\frac{\sqrt{3}Qy}{2}\right), \\
\cos\frac{Qx}{2}\sin\frac{\sqrt{3}Qy}{2},\\
\frac{1}{\sqrt{3}}\left(\cos Qx+2\cos\frac{Qx}{2}\cos\frac{\sqrt{3}Qy}{2}\right)\Bigg],
\end{multline}
which corresponds to the case of ${ \phi_1 = 1 }$ and ${ \theta_1 = 0}$ 
in Eq.~\eqref{eq: SkX}. The wave vector $ Q = |\tilde{D}/\tilde{J}| $ implies a SkX lattice constant of 421~nm. It has the symmetry of a triangular lattice as 
shown in \Fig~\ref{fig: phase_diagram_SkX}(b). 
Therefore Bloch's theorem can be employed to expand the wave function as
\begin{equation}
    \psi\left(\vec r\right)=\me^{\mi\vec k\cdot\vec r}\sum_{n_{1}n_{2}}\me^{\mi(n_{1}\vec b_{1}+n_{2}\vec b_{2})\cdot\vec r}
    \begin{bmatrix}
        u_{n_{1}n_{2}}\\
        v_{n_{1}n_{2}}\\
    \end{bmatrix}
\end{equation}
where $ \vec{k} $ lies in the first Brillouin zone and $ \vec{b}_1 $, $ \vec{b}_2 $ are the two primitive vectors of the reciprocal lattice. 
As in the spin spiral case, Hund's coupling is also chosen to be $ J_{\text{ex}}= 1 $ in units of $\hbar v_{\text{F}} Q$.
We use the cutoff $ |n_1|$ and $|n_2| \le 5 $ to solve the eigenvalue problem, and the band structure is plotted in \Fig~\ref{fig: SkX_surface_state}(a).

In contrast to the spin spiral case, when the Dirac electrons couple to a skyrmion lattice, an energy gap opens at the Fermi level. 
Such a gap opening goes beyond the first order perturbation, because the second term of Eq.~\eqref{eq: electron-spin interaction} does not couple the two degenerate states at the unperturbed Dirac point when $ \langle S_{z}\left(\vec r\right)\rangle=0 $. 
Hence the energy gap is small compared with the value of $ J_{\text{ex}} $.
The spatial distributions of the wave functions of the conduction and valence states at the $ \Gamma $ point are presented in \Fig~\ref{fig: SkX_surface_state}(b). 
It can be seen that the conduction states concentrate around the skyrmion cores, 
while the valence states concentrate around the boundaries between skyrmions. 
It is more comfortable for the Dirac electrons to settle in the regions where 
$ {S_z \approx 0 }$ as expected. The wave-function distributions are similar to the case of topologically trivial electrons coupled to antiferromagnetic skyrmions~\cite{Tome2021}, in which, when the electrons concentrate at the skyrmion centers, they are confined in a ringlike configuration.
The localized electrons charge the skyrmions and have certain overlaps with each other. As a result, one can construct a tight-binding model through the effective Wannier orbitals on a charged SkX and investigate its low-energy properties. 
This physics has been discussed in a very recent work~\cite{Divic2021}.

The coupling between Dirac electrons and the magnetic structure 
breaks the time-reversal symmetry.
Because a gap opens at the Dirac point, the Berry curvatures in general can be nonzero. 
We calculate the Berry curvatures of the valence and conduction bands around the $ \Gamma $ point and present the results in \Fig~\ref{fig: SkX_surface_state}(c). 
It is found that in this region, most of the Berry curvatures distribute near the $ \Gamma $ point and have opposite signs for the conduction and valence bands. 
Because \ce{Bi2Se3} is usually $ n $ doped, there are some electrons at the bottom of the conduction band. 
The nonvanishing Berry curvature of these electrons will bring about the anomalous Hall effect (AHE) on a SkX~\cite{Hamamoto2015}. 
It should be emphasized that the AHE in this system is not quite like the ``conventional'' one, in which the Hall resistivity is empirically proportional to the magnetization, which is zero in our case.

Phenomenologically, the AHE is similar to the topological Hall effect (THE)~\cite{Taguchi2001} commonly found on skyrmion lattices. 
In the THE, the electron experiences an emergent magnetic field that is essentially the
real-space Berry phase of the electron hopping on the magnetic texture
due to the scalar spin chirality, with the electron spin aligned with that of the magnetic moment. 
In our case, the electron is described by a Rashba Hamiltonian with a strong spin-momentum locking, and already has the momentum-space Berry curvature. 
The spin-orbit coupling prevents the electron spin from aligning with the local magnetic moment. Still, the electron spin rotates slightly during the hopping process, and a Berry phase is accumulated for a closed loop in the real space. So this is an example where 
both real-space Berry curvature and momentum-space Berry curvature are present. 
The Berry curvature distribution for both bands is depicted
in \Fig~\ref{fig: SkX_surface_state}(c). 

\section{Discussion and conclusions}\label{sec: Discussions and conclusions}

In our first-principles calculation of the magnetic interactions between the Co adatoms, \ce{Bi2Se3} is assumed to be intrinsic,
i.e., the Fermi surface of the surface electrons is right at the Dirac point. 
In this case, the magnetic interaction mainly comes from direct exchange and superexchange mechanisms. In reality, \ce{Bi2Se3} crystals grown in laboratories are $ n $ doped due to the Se vacancies and antisite defects~\cite{Scanlon2012,Wang2013}. Nevertheless, the Fermi level can be tuned by further doping~\cite{Checkelsky2009,Ren2011}. If the surface electrons have finite Fermi surface, the Ruderman-Kittel-Kasuya-Yosida (RKKY) mechanism will contribute to the magnetic interactions~\cite{Liu2009}. The RKKY interaction was found to be always ferromagnetic when the Fermi level lies near the surface Dirac point~\cite{Liu2009}. Also, the RKKY interaction has a DMI component when considering the spin-orbit coupling~\cite{Ye2010}. Therefore, even for the case of finite doping, the magnetic interaction tensor \eqref{eq: interaction_matrix} remains valid, and the Heisenberg part retains its ferromagnetic nature. We expect only quantitative change in the phase diagram of \Fig~\ref{fig: phase_diagram_SkX}.

The discussion of the topological surface electrons coupling to the magnetic structures can be generalized to other topological materials, especially stoichiometric magnetic topological insulators such as \ce{MnBi2Te4}~\cite{Otrokov2019,Gong2019}, as mentioned in Ref.~[\onlinecite{Paul2021}]. These platforms are easier to fabricate in laboratories.

In conclusion, we calculate the magnetic interactions between cobalt adatoms on the topological insulator \ce{Bi2Se3} (111) surface. The Heisenberg part of the interaction is ferromagnetic. Because of the broken inversion symmetry and strong spin-orbit coupling on the surface, there is also Dzyaloshinskii-Moriya interaction, which twists the spins of Co atoms from a perfectly parallel alignment. We use the Ginzburg-Landau theory to establish the phase diagram of a Co adatom lattice. With the aid of a small external magnetic field, a spin spiral and a skyrmion lattice can be stabilized, besides the ferromagnetic phase. The topological surface state under the influence of a spin spiral, a single skyrmion, and a skyrmion lattice is numerically solved. Chiral conducting modes are found on the domain walls with zero out-of-plane magnetic moment in a spin spiral. Similar chiral modes are also found on the boundary of a single skyrmion. For a skyrmion lattice, a gap opens at the surface Dirac point, leading to the anomalous Hall effect.

\emph{Note added.} Recently, we become aware of two recent works by Paul and Fu~\cite{Paul2021} and by Divic \textit{et al.}~\cite{Divic2021}, in which some similar results are obtained.

\section*{acknowledgments}
We thank Prof. Yayu Wang for comments on the manuscript. The first-principles calculations for this work were performed on TianHe-2. 
We are thankful for the support from the National Supercomputing Center in Guangzhou (NSCC-GZ).
The work is supported by the Ministry of Science and Technology of China 
with Grants No.~2016YFA0301001, No.~2016YFA0300500, and No.~2018YFE0103200, 
by the Shanghai Municipal Science and 
Technology Major Project with Grant No.~2019SHZDZX04, 
and by the Research Grants Council of Hong Kong with 
General Research Fund Grant No.~17306520.

\appendix

\section{Details of the first-principles calculations and the adsorption of cobalt atoms}
\label{app: DFT_details}
In this appendix we describe the model geometry and technical details of the first-principles calculations carried out in this paper. 

For the QL-stacked \ce{Bi2Se3}, we consider one QL slab with experimental lattice parameters~\cite{Mishra1997}, and build a slab model of a $20$-$\mathring{\text{A}}$-thick vacuum layer to investigate the adsorption of cobalt atoms on the surface and magnetic interactions between them. 
The first-principles calculations based on density functional theory (DFT)~\cite{Hohenberg1964,Kohn1965} are performed with plane wave basis sets and pseudopotential method, as implemented in the \textsc{quantum espresso} package~\cite{QE-2009,QE-2017}.
Perdew-Zunger parametrization of the local density approximation~\cite{Perdew1981} is employed for the exchange-correlation functional. 
The projector augmented-wave~\cite{Blochl1994} pseudopotentials in the \textsc{pslibrary}~\cite{DalCorso2014,pslibrary} (version 1.0.0) are adopted. 
The energy cutoff of the plane wave basis set is chosen to be 70~Ry. 
The position of the Co adatom is relaxed by the Broyden-Fletcher-Goldfarb-Shanno (BFGS) quasi-Newton algorithm until the Hellmann-Feynman force is less than 0.001~Ry/bohr. 
The Brillouin zone is sampled by a $ 3 \times 3 \times 1 $ grid in structure relaxations and self-consistent charge density calculations. 
In magnetic interaction calculations, a $ 5 \times 5 \times 1 $ grid is employed. 
Spin-orbit coupling (SOC) is included in all the energy calculations unless explicitly stated otherwise. 

Three typical adsorption positions are calculated, and their energies are compared. 
As marked by colored crosses in Fig.~\ref{fig: lattice_geometry}(b), they are above a hollow (site A), above a Bi atom (site B), and above a Se atom (site C). 
The equilibrium adsorption positions of these sites have different heights relative to the surface Se atomic layer.
Specifically, the site A is $0.24~\mathring{\text{A}}$ below the Se layer, while the site B and the site C are $0.65$ and $2.12~\mathring{\text{A}}$ above the Se layer, respectively.
The adsorption energies on these sites are further computed using a $ 3 \times 3 $ supercell.
It is found that the most stable adsorption site is site A, with energy $ 0.71 $ and $3.14\text{ eV}$ lower than the energies of sites B and C, respectively. 

\section{The determination of spin interaction parameters}
\label{app: spin_interactions}
To calculate the magnetic interactions between two Co adatoms on neighboring A sites, we use a $ 3 \times 3 $ supercell as shown in Fig.~\ref{fig: lattice_geometry}(c).
In addition, we use the energy-mapping method~\cite{Xiang2011,Xiang2013} to determine the interaction parameters in Eq.~\eqref{eq: interaction_matrix} from DFT calculations. 
For example, we would like to calculate the $ J_{xy} $ component. 
Then two magnetic configurations are constructed: (1) $ \vec{S}_{1}=\left(S,0,0\right) $, $ \vec{S}_{2}=\left(0,S,0\right) $; and (2) $ \vec{S}_{1}=\left(S,0,0\right) $, $ \vec{S}_{2}=\left(0,-S,0\right) $. 
We constrain the magnitude and direction of the magnetic moment of the two Co atoms and calculate the total energy of the system, $ E_1 $ and $ E_2 $, by DFT. 
By the spin-spin interaction model in Eq.~\eqref{eq: spin-spin interactions},
\begin{align}
    E_{1} & = E_{0}+J_{xy}S^{2}, \\
    E_{2} & = E_{0}-J_{xy}S^{2},
\end{align}
where $ E_0 $ is the energy of other parts of the system. Then $ J_{xy} $ can be obtained from $ E_1 $ and $ E_2 $,
\begin{equation}
    J_{xy}=\frac{E_{1}+E_{2}}{2}.
\end{equation}
A similar method can be used to calculate $ J_{yx} $; then $ \Gamma_{xy}=\left(J_{xy}+J_{yx}\right)/2 $ (which should be zero by symmetry) and $ D_{z}=\left(J_{xy}-J_{yx}\right)/2 $ can be obtained.

\bibliography{refs}

\end{document}